\documentclass[aps,prl,superscriptaddress,twocolumn,showpacs,preprintnumbers,amsmath,amssymb]{revtex4-2}

\usepackage{graphicx}
\usepackage{dcolumn}
\usepackage{bm}
\usepackage{color}
\usepackage[dvipsnames]{xcolor}

\begin{document}


\title{Wang-MacDonald \textit{d}-wave vortex cores observed in heavily overdoped Bi$_2$Sr$_2$CaCu$_2$O$_{8+\delta}$}

\author{Tim Gazdi\'c}
\affiliation{Department of Quantum Matter Physics, Universit\'e de Gen\`eve,
24 Quai Ernest Ansermet, CH-1211 Geneva 4, Switzerland.}
\author{Ivan Maggio-Aprile}
\affiliation{Department of Quantum Matter Physics, Universit\'e de Gen\`eve,
24 Quai Ernest Ansermet, CH-1211 Geneva 4, Switzerland.}
\author{Genda Gu}
\affiliation{Condensed Matter Physics and Materials Science Department, Brookhaven National Laboratory, Upton, New York 11973, USA}
\author{Christoph Renner}
\email{christoph.renner@unige.ch}
\affiliation{Department of Quantum Matter Physics, Universit\'e de Gen\`eve,
24 Quai Ernest Ansermet, CH-1211 Geneva 4, Switzerland.}

\date{\today}

\begin{abstract}
Low magnetic field scanning tunneling spectroscopy of individual Abrikosov vortices in heavily overdoped Bi$_2$Sr$_2$CaCu$_2$O$_{8+\delta}$ unveils a clear \textit{d}-wave electronic structure of the vortex core, with a zero-bias conductance peak at the vortex center that splits with increasing distance from the core. We show that previously reported unconventional electronic structures, including the low energy checkerboard charge order in the vortex halo and the absence of a zero-bias conductance peak at the vortex center, are direct consequences of short inter-vortex distance and consequent vortex-vortex interactions prevailing in earlier experiments.
\end{abstract}

\maketitle

\textbf{\textit{Introduction}}~\textemdash~High temperature superconductivity (HTS) in copper oxides keeps challenging our understanding. A number of unconventional properties, starting with their high superconducting transition temperature ($T_c$), have sparked sustained theoretical and experimental efforts to explain the underlying electron pairing mechanism \cite{Keimer2015,Fischer2007}. Among the numerous outstanding puzzles, in particular in the extensively studied Bi$_2$Sr$_2$CaCu$_2$O$_{8+\delta}$ (Bi2212), is the electronic structure of the Abrikosov vortex cores. The fundamental excitations bound to magnetic vortices in type-II superconductors carry information about essential properties of the superconducting state. Their proper identification is therefore of prime interest to elucidate the mechanism driving HTS. 

Early scanning tunneling spectroscopy (STS) maps of vortex cores in Bi2212 were neither compatible with the discrete Caroli-de Gennes-Matricon bound states for an \textit{s}-wave superconductor \cite{Caroli1964}, nor with the continuum first calculated by Wang and MacDonald for a \textit{d}-wave superconductor \cite{Wang1995,Franz1998}. Instead of the expected zero bias conductance peak (ZBCP) that splits with increasing distance from the core in the \textit{d}-wave case, they revealed low energy ($E<\Delta_{SC}$) subgap states (SGS) in YBa$_2$Cu$_3$O$_{7-\delta}$ (Y123) \cite{Ivan1995} and a pseudogap-like spectrum in the vortex core of Bi2212 \cite{Renner1998}. Subsequent STS mapping with improved resolution confirmed the presence of SGS in Bi2212 \cite{Hoogenboom2000,Pan2000} and found a $\sim4a_0\times4a_0$ modulation of the local density of states (the checkerboard) spanning the vortex core region \cite{Hoffman2002,Levy2005}. Meanwhile, there was emerging evidence suggesting the SGS and checkerboard were not specific to the vortex core. A non-dispersing $\sim4a_0\times4a_0$ charge density modulation was observed above $T_c$ in the pseudogap (PG) phase of slightly underdoped Bi2212 \cite{Vershinin2004}. Moreover, Hanaguri et al. \cite{Hanaguri2009} found a striking field dependence of the checkerboard in Ca$_{2-x}$Na$_x$CuO$_2$Cl$_2$, suggesting it is a field enhanced quasiparticle interference (QPI) rather than a genuine charge order. Out of phase spatial modulations of the electron-like and hole-like SGS associated with the checkerboard suggest they are indeed rather QPI features \cite{Matsuba2007,Yoshizawa2013}. The wavelength of the periodic charge modulations observed by STS was found to depend on energy below $\Delta_{SC}$ and to be non-dispersing at energies above \cite{Machida2016,Edkins2019}. The dispersion of the low energy features is well described by QPI, while the non-dispersing high energy features have been associated with PG and symmetry breaking charge ordered phases  \cite{Machida2016,Kohsaka2008,Vershinin2004}.

The objective of the present study is to uncover the true electronic structure of a \textit{d}-wave-wave vortex core. A first indication of a Wang-MacDonald vortex core \cite{Wang1995} in a cuprate high temperature superconductor was obtained by STS on Y123 \cite{Berthod2017}. However, in this case, the bare vortex core tunneling spectra were obscured by an unknown spectral signature dominating the superconducting signal. Here we choose to investigate highly overdoped (OD) Bi2212 at low magnetic fields: overdoped, to take advantage of a simpler electronic structure, devoid of any pseudogap and associated electronic phases \cite{Fujita2014}; low field, to increase the vortex spacing and reduce the influence of neighboring vortices, whose screening currents can affect vortex cores over distances set by $\lambda$, the London penetration depth \cite{Berthod2016,Berthod2017}.

\textbf{\textit{Materials and Methods}}~\textemdash~Bi2212 single crystals, grown by the traveling-solvent-floating-zone method, were cleaved at T$\sim100$~K in ultra-high vacuum shortly before their transfer into the STM head at low temperature. The superconducting transition temperature of each sample was measured by magnetic susceptibility. All scanning tunneling microscopy and spectroscopy (STM/STS) experiments were carried out at or below 4.8 K using a commercial SPECS JT Tyto STM \cite{Tyto}, except the temperature dependent STS in the Supplemental Material. We used chemically etched Ir tips, carefully conditioned and characterized on a reconstructed Au(111) single crystal surface. The energy scales in the data refer to the sample bias. A magnetic field in the range of 0-3 Tesla was applied perpendicular to the Bi2212 CuO$_2$ planes. The lowest field of 0.16 Tesla was applied by mounting the sample on a small permanent magnet. The field intensity was inferred from the number of vortices per unit area observed by STS on Bi2212 and on NbSe$_2$ for calibration. Images of the Abrikosov vortex lattice were obtained by mapping the local tunneling conductance at selected biases in a low energy range below the superconducting gap.

\begin{figure}[h]
	\includegraphics[clip=true, width=\columnwidth]{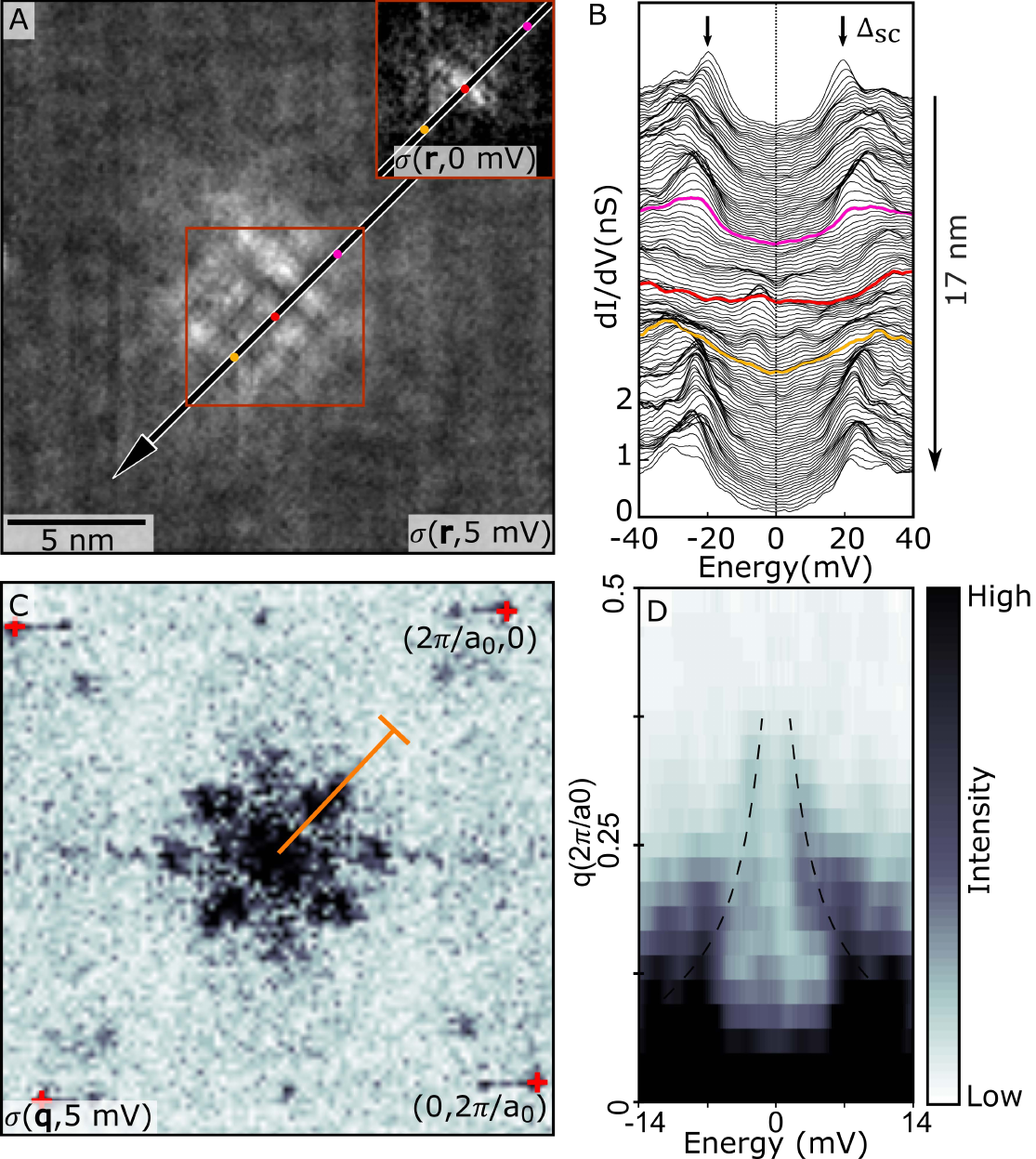}
	\caption{Electronic structure of a vortex core on heavily OD Bi2212 ($T_c\approx52$~K) at 3 Tesla. (\textbf{A}) Low-energy conductance map $\sigma(\textbf{r},\textrm{5 mV})$ showing the $\sim4a_0\times4a_0$ charge density modulation in the vortex halo. The region outlined in red is imaged at 0 mV in the inset to identify the vortex centre (red dot). (\textbf{B}) Differential tunneling conductance spectra measured along the trace through the vortex centre in A (scale corresponds to bottom spectrum, other spectra are offset for clarity). The pink and orange spectra delimit the core region where the SGS develop and the superconducting coherence peaks are suppressed. (\textbf{C}) Fourier transform of A where the lattice peaks at $q_0$ and the periodic modulations at $\sim0.24q_0$ and $\sim0.75q_0$ are clearly resolved. (\textbf{D}) Energy dependence of the Fourier amplitude along the orange trace in C showing the $\sim0.24q_0$ checkerboard dispersion. Dashed lines are guides to the eye.}
	\label{figure:B=3Tvortex}
\end{figure}

\textbf{\textit{High field measurements}}~\textemdash~A single Abrikosov vortex core imaged by STS at 5~mV in a magnetic field of 3 Tesla on heavily OD Bi2212 ($T_c\approx52$~K) is displayed in Fig.\ref{figure:B=3Tvortex}A. Its core region imaged at 0~mV shown in the inset of Fig.\ref{figure:B=3Tvortex}A allows to pinpoint the vortex center highlighted by a red dot. The spatial structure of the 5~mV conductance map looks strikingly similar to earlier STS vortex images obtained on optimally- and slightly under-doped Bi2212. In particular, it shows two periodic patterns extending over the $\sim6$~nm diameter vortex halo. One is reminiscent of the checkerboard reconstruction with $q\approx0.24q_0$ \cite{Hoffman2002,Hanaguri2009}, where  $q_0$ corresponds to the atomic lattice. The other, best resolved in the Fourier transform (FT) shown in Fig.\ref{figure:B=3Tvortex}C, corresponds to a structure similar to the high-energy ladder structure at  q$\approx0.75q_0$ \cite{Kohsaka2007}. Analyzing the FT of conductance images of the same region at different energies, we find the checkerboard pattern to disperse from a characteristic wavevector $q\approx0.375q_0$ at low energy around 3~mV to $q\approx0.24q_0$ at 10~mV (Fig. 1D). Such a dispersion is not consistent with a charge density wave origin of the checkerboard.

Turning to the tunneling spectra, we find the superconducting coherence peaks at $\Delta_{SC}\approx\pm20$~mV greatly suppressed within $\sim 6$~nm of the vortex center along the 17~nm long (100) trace shown in Fig.\ref{figure:B=3Tvortex}B. They further show enhanced low energy SGS developing within the vortex core region delimited by the colored points and spectra in Fig.\ref{figure:B=3Tvortex}A and \ref{figure:B=3Tvortex}B, respectively. These SGS are not compatible with the \textit{d}-wave predictions: their weak amplitude and constant energy position with distance from the core are in sharp contrast to the intense ZBCP and subgap peaks which shift to higher energy with increasing distance from the center predicted for a \textit{d}-wave superconductor \cite{Wang1995,Franz1998}.

\begin{figure}[h]
	\includegraphics[clip=true, width=\columnwidth]{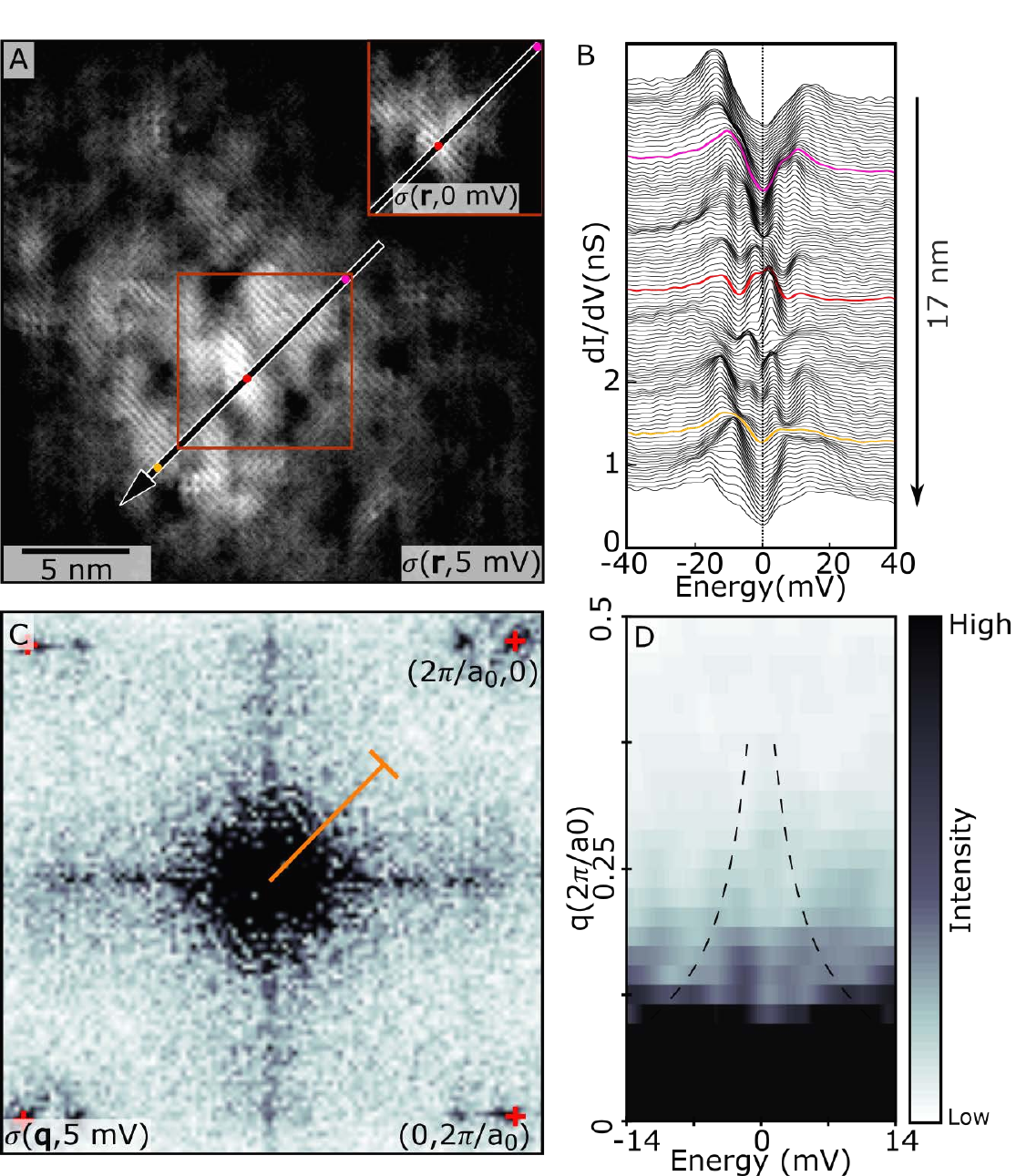}
	\caption{Electronic structure of a vortex core on heavily OD Bi2212 ($T_c\approx52$~K) at 0.16 Tesla. (\textbf{A}) Low-energy conductance map $\sigma(\textbf{r},\textrm{5 mV})$ showing the core of vortex 2 in Fig.\ref{figure:B=0.2Tlattice}. The region outlined in red is imaged at 0 mV in the inset to identify the vortex centre (red dot). (\textbf{B}) Differential tunneling conductance spectra measured along the trace through the vortex centre in A (scale corresponds to bottom spectrum, other spectra are offset for clarity). The pink and orange spectra delimit the core region where the superconducting coherence peaks are suppressed. (\textbf{C}) Fourier transform of A where only the lattice peaks at $q_0$ and features related to the superstructure are resolved. (\textbf{D}) Energy dependence of the Fourier amplitude along the orange trace in C. Note the absence of any characteristic structure. Dashed lines are guides to the eye where dispersive features would arise.}
	\label{figure:B=0.2Tvortex}
\end{figure}

\begin{figure}[h]
	\includegraphics[clip=true, width=\columnwidth]{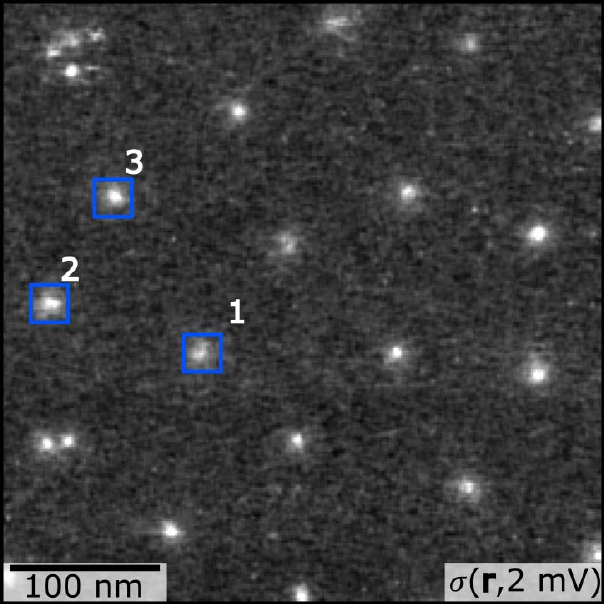}
	\caption{500$\times$500 nm$^2$ STS conductance map of OD Bi2212 ($T_c\approx52$~K) revealing a disordered lattice with a vortex density corresponding to an applied magnetic field of $\sim$0.16 Tesla. Vortex 2 and vortex 3 are further investigated in Fig.\ref{figure:B=0.2Tvortex} and Fig.\ref{figure:SGSdispersion}, respectively.}
	\label{figure:B=0.2Tlattice}
\end{figure}

\textbf{\textit{Low field measurements}}~\textemdash~Compared to the 3 Tesla vortex core structure discussed in Fig.\ref{figure:B=3Tvortex}, a very different picture emerges when reducing the applied magnetic field by over an order of magnitude compared to any previous STS study of a HTS cuprate. The very small coherence length of Bi2212 makes it challenging to ensure that a given feature in an STS map is actually a vortex core and not an impurity or a charge inhomogeneity \cite{SuppInfo}. The individual vortex core shown in Fig.\ref{figure:B=0.2Tvortex} was identified from the large field of view conductance map in Fig.\ref{figure:B=0.2Tlattice}, labelled as vortex 2. This STS map reveals about 20 vortices on a disordered lattice, consistent with an applied magnetic field of 0.16~Tesla. Most remarkably, we observe a systematic presence of a ZBCP at 0.16~Tesla and the disappearance of the vortex checkerboard in vortices sitting in a clean background. This is very different from the vortex cores observed at 3~Tesla, which systematically display the low energy checkerboard and never show a ZBCP.

A single Abrikosov vortex core imaged by STS at 5~mV in a magnetic field of 0.16 Tesla on heavily OD Bi2212 ($T_c\approx52$~K) is shown in Fig.\ref{figure:B=0.2Tvortex}A. It shows no sign of the $q\approx 0.24q_0$ modulation, neither in real space maps (Fig.\ref{figure:B=0.2Tvortex}A) nor in their Fourier transforms (Fig.\ref{figure:B=0.2Tvortex}C). The strong suppression of this modulation at 0.16 Tesla is consistent with its field induced QPI origin \cite{Hanaguri2009}. Most interestingly for the present study, the low field vortices have a significantly different spectroscopic signature compared to 3 Tesla. Besides the suppression of the superconducting coherence peaks observed at all fields, they reveal the characteristic electronic signatures predicted by Wang and MacDonald \cite{Wang1995} for a \textit{d}-wave vortex core (Fig.\ref{figure:B=0.2Tvortex}B): i) a marked conductance peak at or very close to zero bias at the center of the vortex; ii) a splitting of this ZBCP into two subgap conductance peaks (SGCP) that shift away from zero bias with increasing distance from the vortex core.

\textbf{\textit{d-wave core signatures}}~\textemdash~A first conclusion at this stage is that the electronic vortex core structure depends remarkably on the applied magnetic field. Previous STS studies of vortex cores in HTS copper oxides were performed at fields above 2~Tesla. Berthod et al. \cite{Berthod2016,Berthod2017} showed how the screening currents of neighboring vortices at large fields modify the electronic core structure, and in particular the ZBCP and SGCP. These studies further showed that the cores of an irregular vortex lattice can each have a slightly different electronic signature due to their distinct environment. They suggest that in order to access the intrinsic electronic core structure, it is necessary to increase the vortex spacing by reducing the magnetic field. At 0.16 Tesla, we do indeed measure the elusive Wang-MacDonald vortex core structure expected for a \textit{d}-wave superconductor \cite{Wang1995}.

\begin{figure}[h]
	\includegraphics[clip=true, width=\columnwidth]{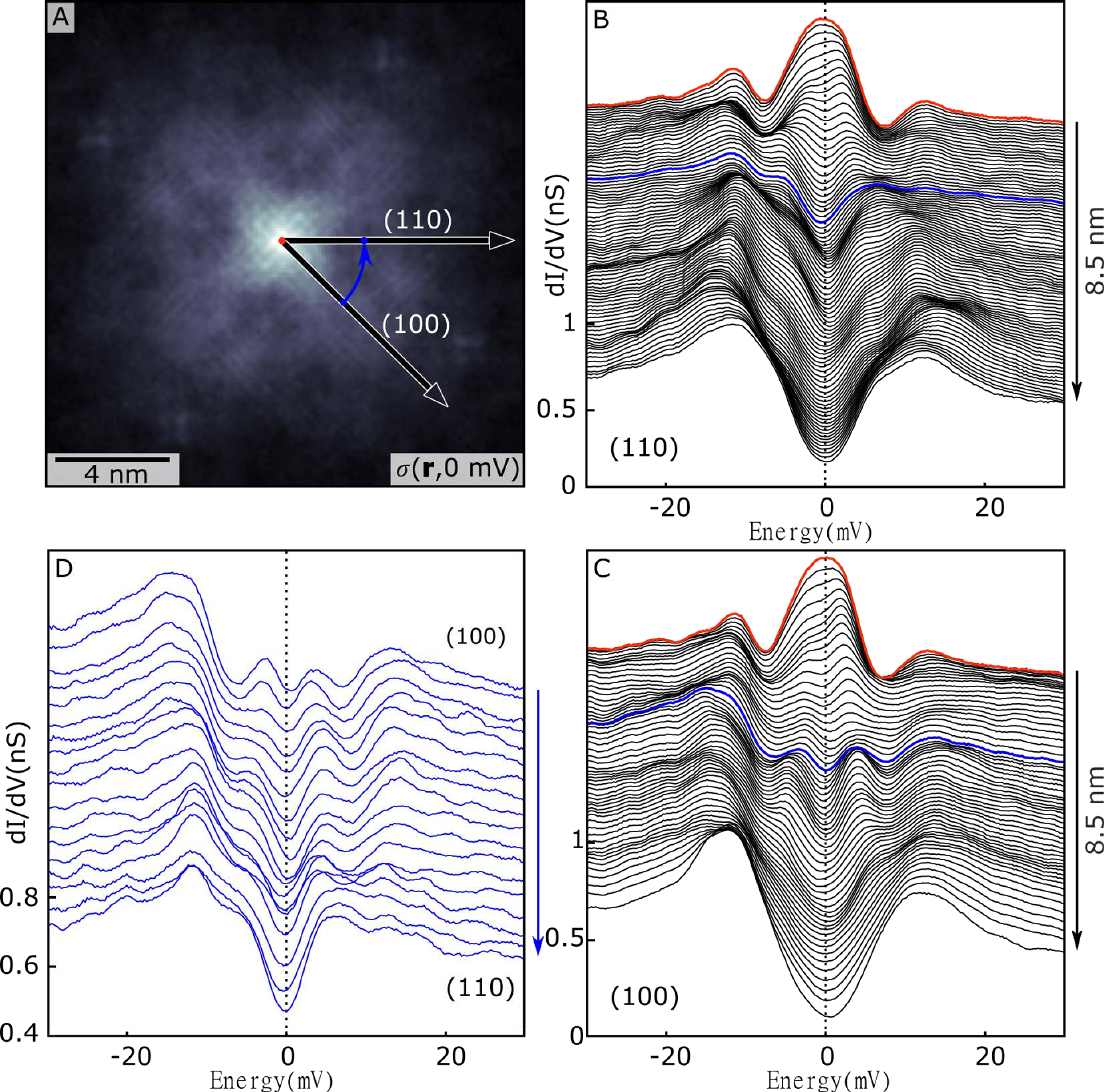}
	\caption{Angular dependence of the low field vortex core structure at 0.16 Tesla. (\textbf{A}) Four-quadrant averaged STS conductance map $\sigma(\textbf{r},\textrm{0 mV})$ of vortex 3 in Fig.\ref{figure:B=0.2Tlattice}. (\textbf{B}) Differential tunneling conductance spectra along the (110) crystallographic direction in A. (\textbf{C}) Differential tunneling conductance spectra along the (100) crystallographic direction in A. (\textbf{D}) Differential tunneling conductance spectra along the arc from (100) to (110) depicted in A. Conductance scales correspond to bottom spectra in panels B, C and D, other spectra are offset for clarity.}
	\label{figure:SGSdispersion}
\end{figure}

The LDOS of an isolated vortex core in an ideal \textit{d}-wave superconductor is four-fold symmetric \cite{Berthod2016,Berthod2017}. We exploit this symmetry to enhance the signal to noise ratio through four quadrant averaging. Such angular averaging enables the resolution of finer electronic features of the cores. Fig.\ref{figure:SGSdispersion}A shows a four-quadrant symmetrized conductance map of the core of vortex 3 measured at 0.16 Tesla in Fig \ref{figure:B=0.2Tlattice}. Line cuts along the nodal (110) and antinodal (100) directions of this averaged data set are shown in Fig.\ref{figure:SGSdispersion}B and \ref{figure:SGSdispersion}C, respectively. Overall, the spatial dependence of the tunneling conductance is similar along both directions, except for a more rapid suppression of the zero-bias conductance (ZBC) along the nodal direction. The suppression of the ZBC evolves smoothly along an arc of constant radius between the nodal and antinodal directions, as seen in Fig.\ref{figure:SGSdispersion}D. These characteristics correspond very closely to the expectations for a \textit{d}-wave vortex core \cite{Berthod2017}. Note that the continuous shift in energy of the SGCP with increasing distance from the core along each direction in Fig.\ref{figure:SGSdispersion} confirms the accurate identification of the vortex center. Indeed, four quadrant averaging around another central point would smear out the above angular dependence. The proper identification of the vortex center is further confirmed by the coherence length extracted from the unsymmetrized vortex core, which is consistent with the known values for this material \cite{SuppInfo}.

Resolving vortex cores at an unprecedented low applied magnetic field in heavily OD Bi2212, is not only instrumental in revealing the \textit{d}-wave electronic vortex core structure, it also allows to delineate extrinsic from genuine vortex features. The present study implies that the checkerboard pattern, the non-shifting SGS and the missing ZBCP are extrinsic features.  They do not signify unconventional Bi2212 vortex cores, but rather result from the interactions of neighboring vortices \cite{Berthod2016,Berthod2017}. The disappearance of the $\sim 4a_0 \times 4a_0$ pattern in the vortex halo at low fields, and the finite dispersion of its corresponding wave vector, indicate that the low energy checkerboard is a field enhanced QPI feature \cite{Hanaguri2009} in agreement with recent measurements close to optimal doping \cite{Machida2016,Edkins2019}. The low field vortex core structure further provides valuable insight into the general electronic structure of Bi2212. For example, the checkerboard and SGS have been linked to the pseudogap in previous studies \citep{Vershinin2004}. Here, we demonstrate that even in the absence of a pseudogap \cite{SuppInfo}, a low energy $\sim 4a_0 \times 4a_0$ modulation and SGS are present when vortex-vortex interactions are significant. 

\textbf{\textit{Summary}}~\textemdash~The key result of the detailed low-field STS spectroscopy discussed here is the experimental demonstration of the vortex core structure predicted by Wang and MacDonald for a \textit{d}-wave superconductor \cite{Wang1995}. This finding removes some of the unusual features previously attributed to \textit{d}-wave vortex cores that have challenged new theories of the high temperature superconducting ground state.

Meanwhile, we find that the low energy electronic structure in high magnetic fields is identical in under- and heavily over-doped Bi2212. Among the consequences of this result is that the low energy electronic signatures are not related to the pseudogap, since the pseudogap is absent in heavily overdoped Bi2212. The remarkable change of the spectroscopic footprint of the vortex cores between 0.16 Tesla and 3 Tesla in heavily overdoped Bi2212 is surprising for such a low energy scale and calls for further investigations.

\section{\label{ACKNOWLEDGMENTS}ACKNOWLEDGMENTS}

\begin{acknowledgments}
We acknowledge C. Berthod for stimulating discussions and for carefully proofreading the manuscript, and A. Scarfato for support with the STS experiments. We thank A. Guipet for technical assistance. Supported by the Swiss National Science Foundation grants 162517 and 182652 (C.R. and T.G.); The work at BNL was supported by the US Department of Energy, oﬃce of Basic Energy Sciences, contract no. de-sc0012704.
\end{acknowledgments}

\end{document}


\title{Supplemental Material for Wang-MacDonald \textit{d}-wave vortex cores observed in heavily overdoped Bi$_2$Sr$_2$CaCu$_2$O$_{8+\delta}$} 

\author{Tim Gazdi\'c}
\affiliation{Department of Quantum Matter Physics, Universit\'e de Gen\`eve,
24 Quai Ernest Ansermet, 1211 Geneva 4, Switzerland.}
\author{Ivan Maggio-Aprile}
\affiliation{Department of Quantum Matter Physics, Universit\'e de Gen\`eve,
24 Quai Ernest Ansermet, CH-1211 Geneva 4, Switzerland.}
\author{Genda Gu}
\affiliation{Condensed Matter Physics and Materials Science Department, Brookhaven National Laboratory, Upton, New York 11973, USA}
\author{Christoph Renner}
\email{christoph.renner@unige.ch}
\affiliation{Department of Quantum Matter Physics, Universit\'e de Gen\`eve,
24 Quai Ernest Ansermet, CH-1211 Geneva 4, Switzerland.}

\date{\today}


\maketitle

\subsection{S1 : Conductance spectra with and without a vortex}

\begin{figure}
	\includegraphics[height=16cm]{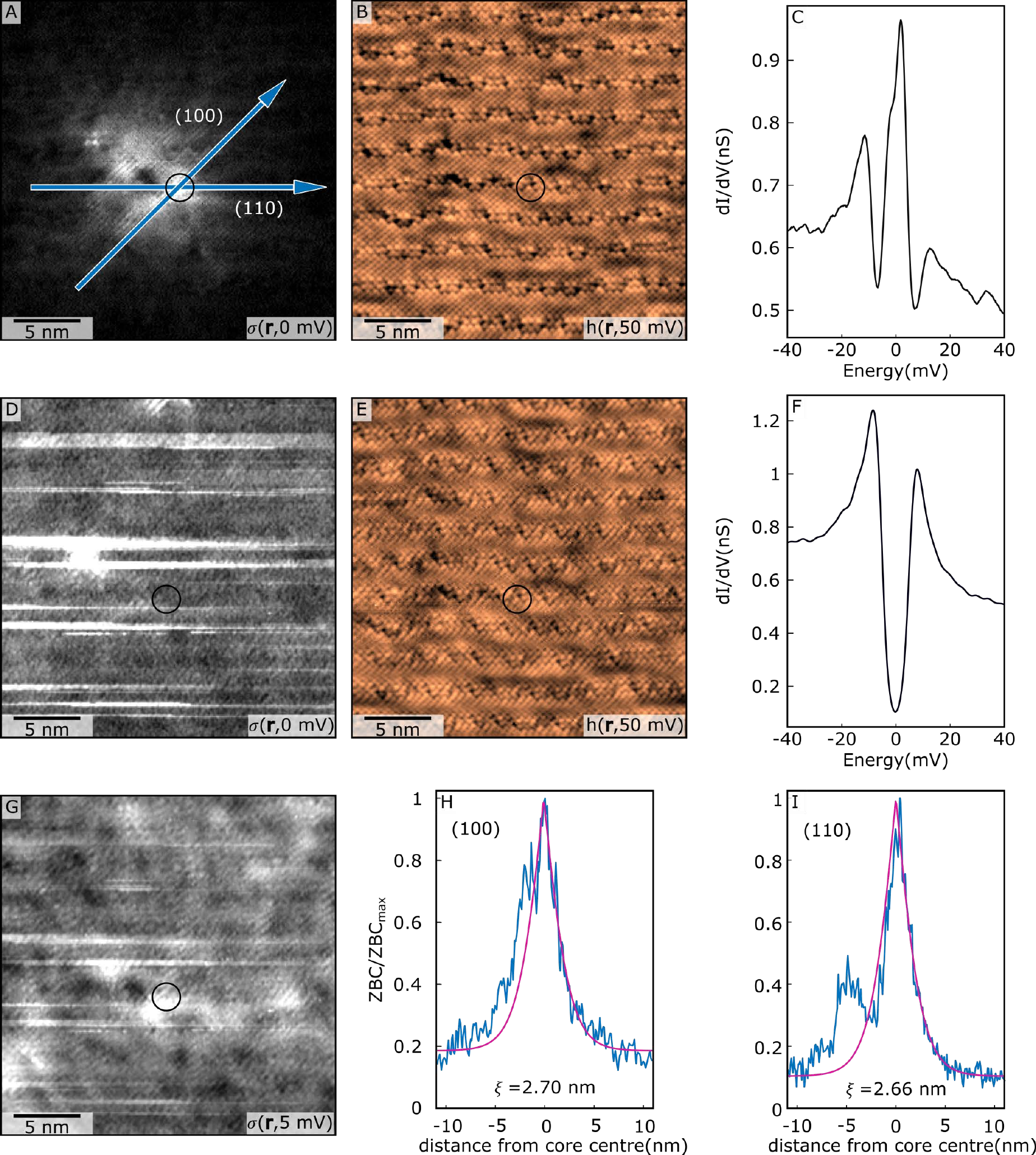}
	\caption{STS and STM images of the exact same region in presence and absence of a vortex on heavily OD Bi2212 ($T_c\approx52$~K). (\textbf{A}) Zero-bias conductance map used to determine the position of the vortex core (circle) and (\textbf{B}) simultaneously acquired high resolution topography in a field of 0.16 Tesla. (\textbf{C}) Differential tunneling conductance spectrum averaged over the vortex core region (circle in A). (\textbf{D}) Zero-bias conductance map and (\textbf{E}) simultaneously acquired high resolution topography in absence of a vortex over the same region as in B.  (\textbf{F}) Differential tunneling conductance spectrum averaged over the region where the vortex core was before cancelling the field (circle). (\textbf{G}) 5 mV conductance map in zero field of the same region as in A and D. (\textbf{H-I}) Normalized zero-bias differential tunneling conductance traces (blue) along the (\textbf{H}) (100) and (\textbf{I}) (110) crystallographic directions through the vortex center highlighted as blue arrows in A. The pink line is a fit of the right side of these profiles with a hyperbolic tangent.}
	
	\label{figure:FigS1}
\end{figure}

Figures~\ref{figure:FigS1}~\textbf{A} to \textbf{F} show STS and STM images of heavily overdoped Bi$_2$Sr$_2$CaCu$_2$O$_{8+\delta}$ (Bi2212), with $T_c\approx52$~K. In order to determine the vortex conductance background, the exact same region was probed in presence and absence of a vortex. From the zero-bias conductance map acquired in a permament magnet field of 0.16 Tesla (Fig.~\ref{figure:FigS1}\textbf{A}), the position of the vortex core center was precisely determined (circle). The simultaneously acquired high resolution topography shown in Fig.~\ref{figure:FigS1}\textbf{B} let us find the exact same region measured in the absence of the vortex - the field of the permanent magnet being cancelled by an applied external field (Figs.~\ref{figure:FigS1}~\textbf{D-E}).  The magnetic field cancellation is not perfect and the bright streaks correspond to an occasional vortex crossing the field of view. The tunneling conductance spectra acquired at the location of the core center reveal a perfect \textit{d}-wave BCS superconducting signature in the absence of the vortex (\textbf{C}), while a clear zero-bias conductance peak is detected when the vortex is present (\textbf{F}). Fig.~\textbf{G} shows a 5 mV conductance map in zero field of the same region as in ~\textbf{A} and ~\textbf{D}, revealing background conductance inhomogeneities which complicate the identification of vortices in Bi2212.

\subsection{S2 : Superconducting coherence length}

Figures~\ref{figure:FigS1}~\textbf{H} and \textbf{I} show normalized zero-bias differential tunneling conductance traces (blue) along the (\textbf{H}) (100) and (\textbf{I}) (110) crystallographic directions through the vortex center highlighted as blue arrows in Fig.~\ref{figure:FigS1}\textbf{A}. Following the procedure described in \cite{Eskildsen2002}, we extract the coherence lengths by fitting the right side of these profiles to a hyperbolic tangent function $f(r)=1-(1-\sigma_0)$tanh$(r/\xi)$ where $\sigma_0$ is the residual conductance far from the center, r is the distance from the center and $\xi$ is the coherence length  (the fits are shown with a pink line in the figure). We find 2.70 nm along (100) and 2.66 nm along (110). The deviations on the left side of the profiles are due to the inhomogeneous background conductance visualized in Fig.~\ref{figure:FigS1}\textbf{G}.

\subsection{S3 : Temperature-dependent measurements}

\begin{figure}
	\includegraphics[height=16cm]{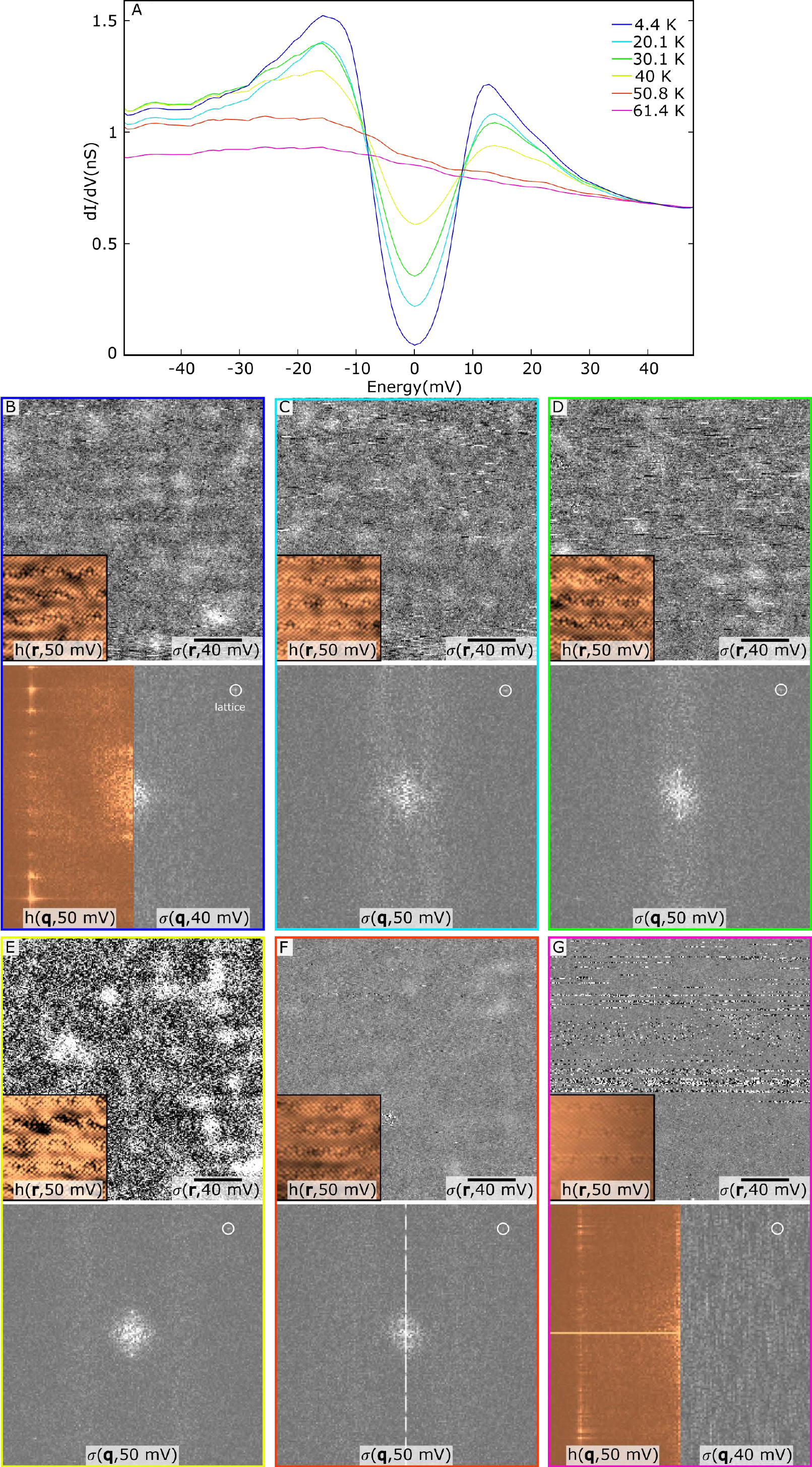}
	\caption{Temperature dependence of the local density of states between 4.4 K and 61.4 K in zero field measured on heavily OD Bi2212 ($T_c\approx52$~K). (\textbf{A}) Differential tunneling conductance spectra show the gap closing above T$_c$ without any sign of a pseudogap. (\textbf{B-G}) STS conductance maps measured as a function of temperature : (\textbf{B})~4.4~K, (\textbf{C})~20.1~K, (\textbf{D})~30.1~K, (\textbf{E})~40~K, (\textbf{F})~50.8~K, (\textbf{G})~61.4~K - at 40~mV. The top panel for each temperature shows the real space conductance, replaced with the simultaneously acquired topographic readout in the inset. The scale bars correspond to 5 nm. The lower panel for each temperature shows the power spectrum (PS) of the corresponding conductance map. In panels B and G, the left side has been replaced by the PS of the topographic image to highlight the lattice peaks which are clearly resolved at all temperatures.}
	\label{figure:FigS2}
\end{figure}

The absence of pseudogap in heavily overdoped Bi2212 ($T_c\approx52$~K)  is evidenced in Fig.~\ref{figure:FigS2}. In Fig.~\ref{figure:FigS2}\textbf{A}, individual tunneling conductance spectra acquired between 4.4~K and 61.4~K  at B~=~0T show that the superconducting gap closes above T$_c$ without any sign of a pseudogap. Similarly, STS maps acquired at the same temperatures (Figs.~\ref{figure:FigS2}~\textbf{B} to \textbf{G}) reveal no pseudogap-related periodic structures found in under- and optimally doped Bi2212 far above the superconducting gap~\cite{Kohsaka2008}.

